\documentclass[11pt, letterpaper]{article}

\usepackage[letterpaper, margin=.75in]{geometry}
\usepackage{float}
\usepackage{multirow}
\usepackage{hyperref}
\usepackage{authblk}
\usepackage{graphicx} 
\usepackage{xcolor}
\usepackage{amsmath, amssymb}
\usepackage{subcaption}
\usepackage{makecell}
\usepackage{multirow}

\newcounter{savenumi}

\newtheorem{theoremfoo}{Theorem}

\newtheorem{openfoo}[theoremfoo]{Question}

\newtheorem{nttn}[theoremfoo]{Notation}

\def\nre.{$n$\/-r.e.}

\newtheorem{factfoo}[theoremfoo]{Fact}

\makeatletter

\def\@makechapterhead#1{ \vspace*{50pt} { \parindent 0pt \raggedright 
 \ifnum \c@secnumdepth >\m@ne \huge\bf \@chapapp{} \thechapter. \par 
 \vskip 20pt \fi \Huge \bf #1\par 
 \nobreak \vskip 40pt } }

\def\@sect#1#2#3#4#5#6[#7]#8{\ifnum #2>\c@secnumdepth
     \def\@svsec{}\else 
     \refstepcounter{#1}\edef\@svsec{\csname the#1\endcsname.\hskip 1em }\fi
     \@tempskipa #5\relax
      \ifdim \@tempskipa>\z@ 
        \begingroup #6\relax
          \@hangfrom{\hskip #3\relax\@svsec}{\interlinepenalty \@M #8\par}
        \endgroup
       \csname #1mark\endcsname{#7}\addcontentsline
         {toc}{#1}{\ifnum #2>\c@secnumdepth \else
                      \protect\numberline{\csname the#1\endcsname}\fi
                    #7}\else
        \def\@svsechd{#6\hskip #3\@svsec #8\csname #1mark\endcsname
                      {#7}\addcontentsline
                           {toc}{#1}{\ifnum #2>\c@secnumdepth \else
                             \protect\numberline{\csname the#1\endcsname}\fi
                       #7}}\fi
     \@xsect{#5}}

\def\@begintheorem#1#2{\it \trivlist \item[\hskip \labelsep{\bf #1\ #2.}]}

\def\@opargbegintheorem#1#2#3{\it \trivlist
      \item[\hskip \labelsep{\bf #1\ #2\ (#3).}]}

\makeatother

\newif\ifshortconferences
\shortconferencesfalse
\newif\ifmediumconferences
\mediumconferencesfalse

\def\ending#1{{\count1=#1\relax
\count2=\count1
\divide\count2 by 100
\multiply\count2 by 100
\advance\count1 by -\count2
\ifnum\count1=11
th%
\else \ifnum\count1=12
th%
\else \ifnum\count1=13
th%
\else 
\count2=\count1
\divide\count1 by 10
\multiply\count1 by 10
\advance\count2 by -\count1
\ifnum\count2=1
st%
\else \ifnum\count2=2
nd%
\else \ifnum\count2=3
rd%
\else th%
\fi\fi\fi\fi\fi\fi
}}

\def\Proceedingsofthe{\ifshortconferences Proc.\else\ifmediumconferences Proc.\else Proceedings of the\fi\fi}

\newcounter{confnum}

\def\conf#1#2{%
\setcounter{confnum}{#2}%
\addtocounter{confnum}{-\csname #1zero\endcsname}%
\ifnum\value{confnum}=1%
\expandafter\ifx\csname #1One\endcsname\relax%
\Proceedingsofthe\ \arabic{confnum}\ending{\value{confnum}}\ \csname #1name\endcsname%
\else \csname #1One\endcsname\fi%
\else%
\Proceedingsofthe\
\arabic{confnum}\ending{\value{confnum}}\ \csname #1name\endcsname\fi}

\def\qsym{\vrule width0.7ex height0.9em depth0ex}

\newif\ifqed\qedtrue

\def\noqed{\global\qedfalse}

\def\qed{\ifqed{\penalty1000\unskip\nobreak\hfil\penalty50
\hskip2em\hbox{}\nobreak\hfil\qsym
\parfillskip=0pt \finalhyphendemerits=0\par\medskip}\fi\global\qedtrue}

\makeatletter
\def\eqnqed{\noqed
	\def\@tempa{equation}
	\ifx\@tempa\@currenvir\def\@eqnnum{\qsym}%
	\addtocounter{equation}{-1}\else%
    \def\@@eqncr{\let\@tempa\relax
    \ifcase\@eqcnt \def\@tempa{& & &}\or \def\@tempa{& &}%
      \else \def\@tempa{&}\fi
     \@tempa {\def\@eqnnum{{\qsym}}\@eqnnum}
     \global\@eqnswtrue\global\@eqcnt\z@\cr}\fi}

\def\eqnlabel#1#2{\if@filesw {\let\thepage\relax%
   \def\protect{\noexpand\noexpand\noexpand}%
   \edef\@tempa{\write\@auxout{\string
      \newlabel{#2}{{{#1}}{\thepage}}}}%
   \expandafter}\@tempa%
   \if@nobreak \ifvmode\nobreak\fi\fi\fi%
	\def\@tempa{equation}
	\ifx\@tempa\@currenvir\def\theequation{{#1}}%
	\addtocounter{equation}{-1}\else%
    \def\@@eqncr{\let\@tempa\relax
    \ifcase\@eqcnt \def\@tempa{& & &}\or \def\@tempa{& &}%
      \else \def\@tempa{&}\fi
     \@tempa {\def\@eqnnum{{#1}}\@eqnnum}
     \global\@eqnswtrue\global\@eqcnt\z@\cr}\fi}

\makeatother

\def\QED{\qed}

\makeatother

\usepackage{amsthm}
\newtheorem{theorem}{Theorem}
\newtheorem{lemma}{Lemma}

\newtheorem{definition}{Definition}

\newcommand{\parae}[1]{\vspace{.2cm}\noindent\emph{#1}}
\newcommand{\parab}[1]{\vspace{.2cm}\noindent\textbf{#1}}

\title{Computing Threshold Circuits with Bimolecular Void Reactions in Step Chemical Reaction Networks\thanks{This research was supported in part by National Science Foundation Grant CCF-2329918.}}

\author[1]{Rachel Anderson}
\author[1]{Bin Fu}
\author[1]{Aiden Massie}
\author[1]{Gourab Mukhopadhyay}
\author[1]{Adrian Salinas}
\author[1]{Robert Schweller}
\author[2]{Evan Tomai}
\author[1]{Tim Wylie}

\affil[1]{University of Texas Rio Grande Valley}
\affil[1]{University of Texas Dallas}
\date{}
\begin{document}
\maketitle

\begin{abstract}


Step Chemical Reaction Networks (step CRNs) are an augmentation of the Chemical Reaction Network (CRN) model where additional species may be introduced to the system in a sequence of ``steps.'' We study step CRN systems using a weak subset of reaction rules, \emph{void} rules, in which molecular species can only be deleted. We demonstrate that step CRNs with only void rules of size (2,0) can simulate threshold formulas (TFs) under linear resources. These limited systems can also simulate threshold \emph{circuits} (TCs) by modifying the volume of the system to be exponential. We then prove a matching exponential lower bound on the required volume for simulating threshold circuits in a step CRN with (2,0)-size rules under a restricted \emph{gate-wise} simulation, thus showing our construction is optimal for simulating circuits in this way.
\end{abstract}

\section{Introduction}

Chemical Reaction Networks (CRNs) are a well-established model of chemistry. In this model, chemical interactions are modeled as molecular species that react to create products according to a set of reaction rules. CRNs have been extensively studied since their standard formulation in the 1960s \cite{Aris:1965:ARMA,Aris:1968:ARMA}. Several equivalent models were also introduced around the same time with Vector Addition Systems (VASs) \cite{Karp:1969:JCSS} and Petri-nets \cite{Petri:1962:PHD}. Further, Population Protocols \cite{Angluin:2006:DC} are a restricted form of the model focused on bimolecular reactions. 

\parae{Step CRNs.} These models all assume a discrete starting number of species or elements and reaction rules that dictate how they can interact. Thus, any change in species numbers is only through these interactions. Motivated by standard laboratory procedures where additional chemical species may be added to an initial container of species after a set of reactions has passed, we utilize an extension to the CRN model known as the Step Chemical Reaction Network model (step CRN) first introduced in \cite{anderson2024computing}. The step CRN model adds a sequence of discrete steps where additional species can be added to the existing CRN configuration after waiting for all possible reaction rules to occur in the system.

\parae{Bimolecular Void Rules.} In this paper, we use the terms \emph{reaction} and \emph{rule} interchangeably to denote a reaction rule. We focus on step CRN systems that include only \emph{bimolecular void rules} ($(2,0)$ rules) which are CRN rules that have two reactants and no products, and thus can only delete existing species.  Void rules of size $(2,0)$ and $(3,0)$ were first studied within the standard CRN model in the context of the reachability problem~\cite{Alaniz:2022:ARXIV}.  While standard CRN rules are powerful thanks to their ability to replace, delete or create new species, (2,0) rules have extremely limited power to compute even simple functions in a standard CRN~\cite{anderson2024computing}.  In contrast, we show $(2,0)$ void rules become capable of computing Threshold Formulas (TF) and Threshold Circuits (TC) in the step CRN model.
Threshold circuits are a computationally universal class of Boolean circuits that have practical application in deep learning, and consist of any Boolean circuit made of AND, OR, NOT, and MAJORITY gates.



\parae{Computation in Chemical Reaction Networks.} Computation within Chemical Reaction Networks is a well-studied topic. Within stochastic Chemical Reaction Networks with some possibility for error, the systems are Turing-complete \cite{soloveichik2008computation}. In contrast, error-free stochastic CRNs are capable of computing semilinear functions \cite{angluin2007computational,chen2014deterministic}.
Further, molecules themselves have long been studied as a method of information storage and Boolean logic computation. In particular, CRNs and similar models have been extensively studied in these areas \cite{arkin1994computational,beiki2018real,cardelli2018chemical,Cook:2009:AB,ellis2019robust,jiang2013digital,lin2020mining,qian2011scaling,Qian_Winfree_2011}. Logic gates such as AND \cite{cardelli2018chemical,dalchau2015probabilistic,magri2009fluorescent,Qian_Winfree_2011,thachuk2015leakless,Xiao_Zhang_Zhang_Chen_Shi_2020}, OR \cite{cardelli2018chemical,dalchau2015probabilistic,Qian_Winfree_2011,thachuk2015leakless}, NOT \cite{cardelli2018chemical}, XOR \cite{cardelli2018chemical,Xiao_Zhang_Zhang_Chen_Shi_2020}, NAND \cite{cardelli2018chemical,Cook:2009:AB,ellis2019robust,winfree2019chemical}, NOR \cite{cardelli2018chemical}, Parity \cite{eshra2013odd,Fan_Fan_Wang_Dong_2018,fan2022engineering} and Majority \cite{angluin2008simple,cardelli2012cell,Mailloux_Guz_Zakharchenko_Minko_Katz_2014} have also been explored.


\begin{table}[t]
    \centering\renewcommand{\arraystretch}{1.3}
    \begin{tabular}{| c | c | c  | c | c | c | c |}\hline
    \multicolumn{7}{|c|}{\textbf{Function Computation}}\\ \hline\hline
     \textbf{Rules} & \textbf{Species} & \textbf{Steps} & 
     \textbf{Volume} & \textbf{Simulation} & \textbf{Family} & \textbf{Ref} \\ \hline
     
     $(2,0)$ & $O(G)$ & $O(D)$ & $O(G)$ & Gate-Wise & \textbf{TF} Formulas & Theorem~\ref{(2,0)_formula} \\ \hline
     $(2,0)$ & $O(G)$ & $O(D)$ & $O(G {F_{out}}^D)$ & Gate-Wise & \textbf{TC} Circuits & Theorem~\ref{(2,0)_circuit} \\ \hline
     
     $(2,0)$ & - & - & $2^{\Omega (D)}$ & Gate-Wise & \textbf{TC} Circuits & Theorem~\ref{(2,0)-lower-thm} \\
     \hline 
    \end{tabular}
\caption{Results in the paper related to computing circuits with $(2,0)$ void rules in step CRN systems. $D$ is the circuit's depth, $G$ is the number of gates in a circuit, and $F_{out}$ is the maximum fan-out of a gate in the circuit.}\label{tab:contributions}
\end{table}

\parab{Our Contributions.} 
Table \ref{tab:contributions} gives an overview of our results, and the paper is formatted to introduce the general techniques and then expand into the necessary details to prove these results.
Section \ref{2_0_gates} shows how a step CRN, using (2,0) void rules, computes individual logic gates. We then show how these gates can be combined to build a general construction of threshold formulas in Section \ref{2_0_computation}. Theorem \ref{(2,0)_formula} shows how a step CRN with only (2, 0) rules is capable of computing threshold formulas with $O(G)$ species, $O(D)$ steps, and $O(G)$ volume, where $G$ is the number of gates in a circuit and $D$ is the depth of a circuit. 

In Section \ref{2_0_circuits}, we modify this construction to compute threshold circuits with $O(G)$ species, $O(D)$ steps, and $O(G {F_{out}}^D)$ volume, where $F_{out}$ is the maximum fan-out of the circuit. Finally, in Section \ref{2_0_lower_bound} we show that the volume lower bound for simulating a circuit using \emph{gate-wise} simulation in a step CRN with (2,0) rules is $2^{\Omega(D)}$.  This lower bound is of note in that it shows the exponential volume utilized by the positive result is needed for this style of computation, and it shows a provable change in power from the polynomial volume achievable with $(3,0)$ void rules~\cite{anderson2024computing}.
\section{Preliminaries}

\subsection{Chemical Reaction Networks}

\parae{Basics.}
Let $\Lambda= \{\lambda_1, \lambda_2, \ldots, \lambda_{|\Lambda|}\}$ denote some ordered alphabet of \emph{species}. A configuration over $\Lambda$ is a length-$|\Lambda|$ vector of non-negative integers that denotes the number of copies of each present species.  A \emph{rule} or \emph{reaction} has two multisets, the first containing one or more \emph{reactant} (species), used for creating resulting \emph{product} (species) contained in the second multiset. Each rule is represented as an ordered pair of configuration vectors $R=(R_r, R_p)$. $R_r$  contains the minimum counts of each reactant species necessary for reaction $R$ to occur, where reactant species are either \emph{consumed} by the rule in some count or leveraged as \emph{catalysts} (not consumed); in some cases a combination of the two. The product vector $R_p$ has the count of each species \emph{produced} by the \emph{application} of rule $R$, effectively replacing vector $R_r$.  The species corresponding to the non-zero elements of $R_r$ and $R_p$ are termed \emph{reactants} and \emph{products} of $R$, respectively.

The \emph{application} vector of $R$ is $R_a = R_p - R_r$, which shows the net change in species counts after applying rule $R$ once. For a configuration $C$ and rule $R$, we say $R$ is applicable to $C$ if $C[i] \geq R_r[i]$ for all $1\leq i\leq |\Lambda|$, and we define the \emph{application} of $R$ to $C$ as the configuration $C' = C + R_a$. For a set of rules $\Gamma$, a configuration $C$, and rule $R\in \Gamma$ applicable to $C$ that produces $C' = C + R_a$, we say $C \rightarrow^1_\Gamma C'$, a relation denoting that $C$ can transition to $C'$ by way of a single rule application from $\Gamma$.
We further use the notation $C\rightsquigarrow_\Gamma C'$ to signify the transitive closure of $\rightarrow^1_\Gamma$ and say $C'$ is \emph{reachable} from $C$ under $\Gamma$, i.e., $C'$ can be reached by applying a sequence of applicable rules from $\Gamma$ to initial configuration $C$. Here, we use the following notation to depict a rule $R=(R_r, R_p)$: $ \sum_{i=1}^{|\Lambda|} R_r[i]s_i \rightarrow \sum_{i=1}^{|\Lambda|} R_p[i]s_i$.

Using this notation, a rule turning two copies of species $H$ and one copy of species $O$ into one copy of species $W$ would be written as $2H + O \rightarrow W$.

\begin{definition}[Discrete Chemical Reaction Network]  
    A discrete chemical reaction network (CRN) is an ordered pair $(\Lambda, \Gamma)$ where $\Lambda$ is an ordered alphabet of species, and $\Gamma$ is a set of rules over $\Lambda$. 
\end{definition}

We denote the set of reachable configurations for a CRN $(\Lambda,\Gamma)$ from initial configuration $I$ as $\text{REACH}_{I,\Lambda,\Gamma}$. A configuration is called \emph{terminal} with respect to a CRN $(\Lambda, \Gamma)$ if no rule $R \in \Gamma$ can be applied to it. We define the subset of reachable configurations that are terminal as $\text{TERM}_{I,\Lambda,\Gamma}$. For an initial configuration $I$, a CRN $(\Lambda, \Gamma)$ is said to be \emph{bounded} if a terminal configuration is guaranteed to be reached within some finite number of rule applications starting from configuration $I$.

\subsection{Void Rules}

\begin{definition}[Void and Autogenesis rules]
    A rule $R=(R_r, R_p)$ is a \emph{void} rule if $R_a = R_p - R_r$ has no positive entries and at least one negative entry.  A rule is an \emph{autogenesis} rule if $R_a$ has no negative values and at least one positive value.  If the reactants and products of a rule are each multisets, a void rule is a rule whose product multiset is a strict submultiset of the reactants, and an autogenesis rule one where the reactants are a strict submultiset of the products. There are two classes of void rules, \emph{catalytic} and \emph{true void}. In catalytic void rules, one or more reactants remain after the rule is applied. In true void rules, such as $(2,0)$ and $(3,0)$ rules, there are no products remaining.  
\end{definition}

\begin{definition}
    The \emph{size/volume} of a configuration vector $C$ is $\verb"volume"(C) = \sum C[i]$. 
\end{definition}

\begin{figure}[t]
    \centering
    \includegraphics[width=1.\textwidth]{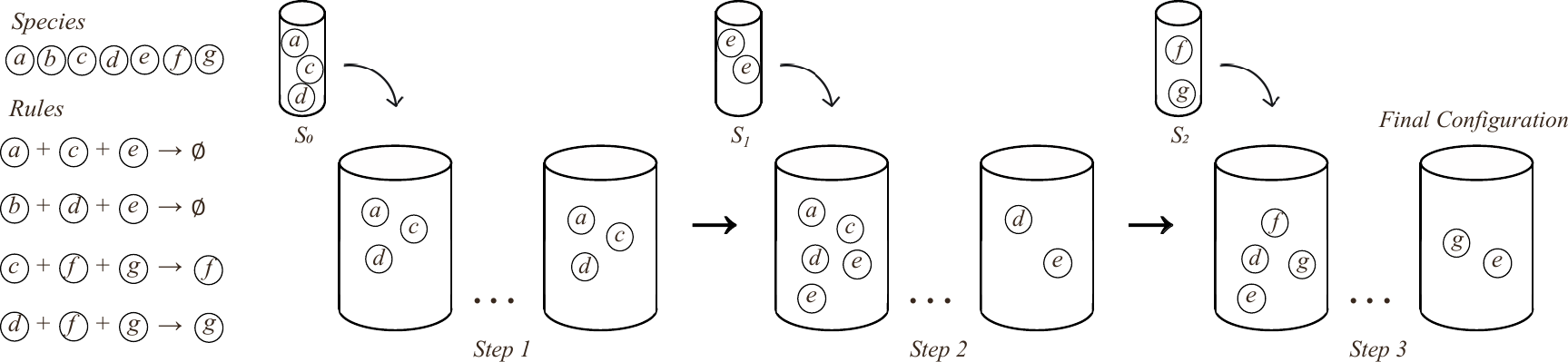}
    \vspace{-.2cm}
    \caption{An example step CRN system. The test tubes show the species added at each step and the system with those elements added. The CRN species and void rule-set are shown on the left.
    \vspace*{-.5cm}}
    \label{fig:step crn species and rules}
\end{figure}

\begin{definition}[size-$(i,j)$ rules]
A rule $R=(R_r, R_p)$ is said to be a size-$(i,j)$ rule if $(i,j) = (\verb"volume"(R_r),$ $\verb"volume"(R_p))$. A reaction is 
bimolecular if $i = 2$.
\end{definition}


\subsection{Step Chemical Reaction Networks} 
A step CRN is an augmentation of a basic CRN in which additional copies of some system species are added after each of a sequence of steps.  Formally, a step CRN of $k$ steps is a ordered pair $( (\Lambda,\Gamma), (s_0, s_1, s_2, \ldots s_{k-1}))$, where the first element of the pair is a normal CRN $(\Lambda,\Gamma)$, and the second is a sequence of length-$|\Lambda|$ vectors of non-negative integers denoting how many copies of each species type to add after each step.

Given a step CRN, we define the set of reachable configurations after each sequential step.  To start off, let $\text{REACH}_1$ be the set of reachable configurations of $(\Lambda, \Gamma)$ with initial configuration $s_0$, which we refer to as the set of configurations reachable \emph{after step 1}.  Let $\text{TERM}_1$ be the subset of configurations in $\text{REACH}_1$ that are terminal.  Note that after just a single step we have a normal CRN, i.e., 1-step CRNs are just normal CRNs with initial configuration $s_0$.  For the second step, we consider any configuration in $\text{TERM}_1$ combined with $s_1$ as a possible starting configuration and 
define $REACH_2$ to be the union of all reachable configurations from each possible starting configuration attained by adding $s_1$ to a configuration in $TERM_1$.
We then define $\text{TERM}_2$ as the subset of configurations in $\text{REACH}_2$ that are terminal.  
Similarly, define $REACH_i$ to be the union of all reachable sets attained by using initial configuration $c_{i - 1}$ at step $s_{i - 1}$ plus any element of $TERM_{i-1}$, and let $TERM_i$ denote the subset of these configurations that are terminal.


The set of reachable configurations for a $k$-step CRN is the set $\text{REACH}_{k}$, and the set of terminal configurations is $\text{TERM}_k$. A classical CRN can be represented as a step CRN with $k=1$ steps and an initial configuration of $I=s_0$.

\subsection{Computing Functions in Step CRNs}\label{def-step-CRN-section}


Here, we define what it means for a step CRN to compute a function $f(x_1,\ldots, x_n) = (y_1,\ldots, y_m)$ that maps $n$-bit strings to $m$-bit strings.  For each input bit, we denote two separate species types, one representing bit 0, and the other bit 1.  An input configuration to represent a desired $n$-bit input string is constructed by selecting to add copies of \emph{either} the 0 species or the 1 species for each bit in the target bit-string.  Similarly, each output bit has two species representatives (for 0 and 1), and we say the step CRN computes function $f$ if for any given $n$-bit input $x_1,\ldots, x_n$, the system obtained by adding the species for the string $x_1,\ldots, x_n$ to the initial configuration of this system in step 1 results in a final configuration whose output species encode the string $f(x_1,\ldots, x_n)$. Note that for a fixed function $f$, the species $s_i$ added at each step are fixed to disallow outside computation.  We now provide a more detailed formalization of this concept.

\parae{Input-Strict Step CRN Computing.} Given a Boolean function $f(x_1,\ldots , x_n) = (y_1,\ldots , y_m)$ that maps a string of $n$ bits 
to a string of $m$ bits, 
we define the computation of $f$ with a step CRN. An input-strict step CRN computer is a tuple $C_s = (S,X,Y)$ where $S = ( (\Lambda,\Gamma), (s_0, s_1, \ldots , s_{k-1}))$ is a step CRN,
and $X=( (x^F_1,x^T_1),\ldots ,(x^F_n,x^T_n))$ and $Y=( (y^F_1,y^T_1),\ldots , (y^F_m,y^T_m))$ are sequences of ordered-pairs with each $x^F_i,x^T_i,y^F_j,y^T_j \in \Lambda$. Given an $n$-input bit string $b=b_1,\ldots , b_n$, configuration $X(b)$ is defined as the configuration over $\Lambda$ obtained by including one copy of $x^F_i$ only if $b_i=0$ and one copy of $x^T_i$ only if $b_i=1$ for each bit $b_i$. 
We consider this representation of the input to be strict, as opposed to allowing multiple copies of each input bit species.
The corresponding step CRN $(\Lambda, \Gamma, (s_0 + X(b), \ldots , s_{k-1}))$ is obtained by adding $X(b)$ to $s_0$ in the first step, which conceptually represents the system programmed with 
input $b$.

An input-strict step CRN computer \emph{computes} function $f$ if, for all $n$-bit strings $b$ and for all terminal configurations of $(\Lambda, \Gamma, (s_0 + X(b), \ldots , s_{k-1}))$, the terminal configuration contains at least 1 copy of $y^F_j$ and 0 copies of $y^T_j$ if the $j^{th}$ bit of $f(b)$ is 0, and at least 1 copy of $y^T_j$ and 0 copies of $y^F_j$ if the $j^{th}$ bit of $f(b)$ is 1, for all $j$ from $1$ to $m$.

We use the term \emph{strict} to denote requiring exactly one copy of each bit species.  While previous work has focused on strict computation~\cite{anderson2024computing}, the focus here is a relaxation of this requirement in which multiple copies of each input species are permitted.

\parae{Multiple Input Relaxation.}  In this paper, we focus on a relaxation to strict computing that allows for multiple copies of each input bit species, i.e., we modify the strict definition by allowing some number greater or equal to 1 of each $x^F_i$ or $x^T_i$ species in the initial configuration 
(but still requiring 0 copies of the alternate choice species).  A system that computes a function under this relaxation is said to be \emph{multiple input relaxed}.


\parae{Gate-Wise Simulation.} In this paper, we utilize a method of simulation we term \emph{gate-wise} simulation, where the output of each gate is represented by a unique species and the gates are computed in the order of their depth level. In other words, multiple gates cannot use the exact same species to represent their output and a gate can only be computed once all gates in the previous depth levels have been computed. We use this method when simulating formulas and circuits with a step CRN in Sections \ref{2_0_formulas} and \ref{2_0_circuits}. 
A formal definition of gate-wise simulation is provided in Section \ref{2_0_lower_bound}.

\subsection{Boolean and Threshold Circuits}

A Boolean circuit on $n$ variables $x_1, x_2, \ldots, x_n$ is a directed, acyclic multi-graph.  The vertices of the graph are referred to as \emph{gates}.  The in-degree and out-degree of a gate are called the \emph{fan-in} and \emph{fan-out} of the gate, respectively.  The fan-in 0 gates  (\emph{source} gates) are labeled from $x_1, x_2, \ldots, x_n$, or from the constants $0$ or $1$.  Each non-source gate is labeled with a function name, such as AND, OR, or NOT.  Fan-out 0 gates may or may not be labeled as output gates.  Given an assignment of Boolean values to variables $x_1, x_2, \ldots, x_n$, each gate in the circuit can be assigned a value by first assigning all source vertices the value matching the labeled constant or labeled variable value, and subsequently assigning each gate the value computed by its labeled function on the values of its children.  Given a fixed ordering on the output gates, the sequence of bits assigned to the output gates denotes the value computed by the circuit on the given input.  

The \emph{depth} of a circuit is the longest path from a source vertex to an output vertex.  A circuit is a called a \emph{formula} if all non-source vertices have fan-out 1, and there is exactly 1 output gate. 
Here, we focus on circuits that consist of AND, OR, NOT, and MAJORITY gates with arbitrary fan-in. We refer to circuits that use these gates as \textit{Threshold Circuits} (TC).

\parae{Notation.}  When discussing a Boolean circuit, the following variables are used to denote the properties of the circuit: $G$ denotes the number of gates in the circuit, $D$ the circuit's depth, and $F_{out}$ the maximum fan-out of any gate in the circuit.
\section{Computation of Threshold Formulas with (2, 0) Rules}\label{2_0_formulas}

Section \ref{2_0_bits} and Section \ref{2_0_gates} introduce how a step CRN using only (2,0) rules represents bits and 
logic gates of a Threshold Formula $(TFs)$, respectively. 
An example construction of a formula is provided in Section \ref{2_0_example}. Section \ref{2_0_computation} then shows the general construction of building TFs, and we prove how the system computes TFs with $O(G)$ species, $O(D)$ steps, and $O(G)$ volume in Theorem \ref{(2,0)_formula}.

\begin{figure}[t]
    \centering
    \begin{subfigure}[t] {0.35\textwidth}
        \centering
        \includegraphics[width=.8\textwidth]{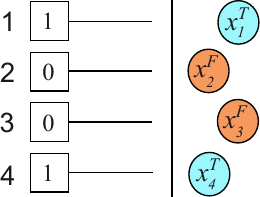}
        \caption{}
        \label{fig:bitrep}
    \end{subfigure}
    \begin{subfigure}[t] {0.45\textwidth}
        \centering
        \includegraphics[width=0.85\textwidth]{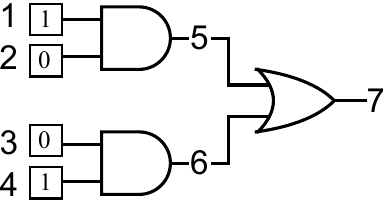}
        \caption{}
        \label{fig:indexingformula}
    \end{subfigure}
    \caption{(a) The input bits of a threshold formula 
    and their representation as species. (b) An indexed threshold formula with the input species shown in Figure \ref{fig:bitrep}.}
    \label{fig:xyz}
\end{figure}

\subsection{Bit Representation}\label{2_0_bits}
Here, we show how the bits of a TF are represented in our model. We first demonstrate a system for indexing the TF's wires 
before introducing the species used to represent bits.

\parae{Indexing.} Every wire of the TF has a unique numerical index. The input and output bits that traverse these wires also use the wire's index. If a wire fans out, the fanned-out wires share the same index as the original. This indexing ensures that the bit species only use the rules that compute their respective gates. Note that gates may also be denoted with an index, where its index is that of its output wire. 



\parae{Bits.} Every input bit of a binary gate is represented by the species $x_n^b$, where $n \in \mathbb{N}$ and $b \in \{ T, F \}$. $n$ represents the bit's index and $b$ represents its truth value. An example of these species is shown in Figure \ref{fig:bitrep}.
Every output bit of a binary gate is represented by the species $y_n^b$ or $y_{j \rightarrow i}^b$, where $j$ represents the input bit's ($x^b$) index and $i$ represents the output bit's ($y^b$) index.


\subsection{Logic Gate Computation}\label{2_0_gates}
We now show how the logic gates of a TF are computed. Let $f_i^{in}$ be the set of all the indexes of the inputs fanning into a gate at index $i$. 




\begin{figure}[t]
    \centering
    \begin{subfigure}[t] {0.45\textwidth}
        \centering
        \includegraphics[width=.9\textwidth]{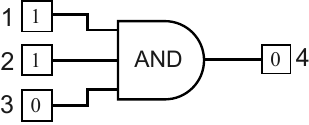}
        \caption{}
        \label{fig:andgate}
    \end{subfigure}
    \begin{subfigure}[t] {0.45\textwidth}
        \centering
        \includegraphics[width=.7\textwidth]{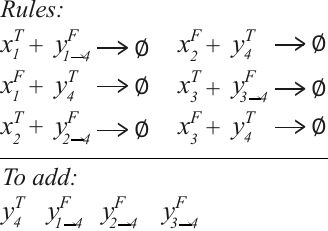}
        \caption{}
        \label{fig:andgaterules}
    \end{subfigure}
        \bigskip 
    \begin{subfigure}[t] {0.45\textwidth}
        \centering
        \includegraphics[width=1.\textwidth]{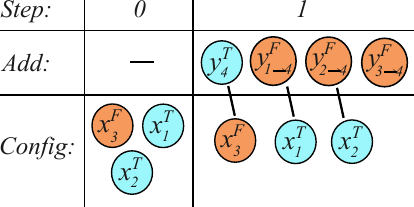}
        \caption{}
        \label{fig:andgatesteps}
    \end{subfigure}
    \begin{subfigure}[t] {0.45\textwidth}
        \centering
        \includegraphics[width=1.\textwidth]{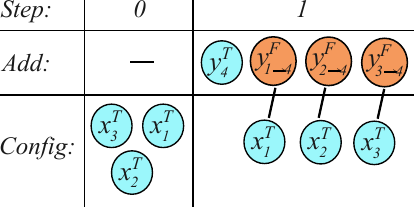}
        \caption{}
        \label{fig:andgatesteps2}
    \end{subfigure}
    \label{andgatefigure}
    \caption{(a) A threshold formula consisting of a single three-input AND gate. (b) Reaction rules and added species for the step CRN that compute the formula in \ref{fig:andgate}. (c) The step CRN computing the formula in Figure \ref{fig:andgate}. The black lines connecting species represents a reaction applied to them. (d) The step CRN computing the circuit in Figure \ref{fig:andgate}, but with three true inputs. 
    }\label{fig:blah}
\end{figure}


\parae{AND Gate.} To compute an AND gate such as the one shown in Figure \ref{fig:andgate}, a single true output species $y_i^T$ and $|f_i^{in}|$ copies of the false output species $y_{j \rightarrow i}^F$ are added in. In one step, the true input and false output species delete each other. Additionally, if at least one false input species exists, it deletes the lone true output species along with itself, guaranteeing a false output as shown in the example steps in Figure \ref{fig:andgatesteps}. The only output species remaining after the step are those whose truth value matches the intended output.




\parae{AND Gate Example.} Consider an AND gate with index 4 and a fan-in of three; the first two inputs are true and the last is false. For computing this gate with our model, the system's initial configuration consists of $x_1^T$, $x_2^T$, and $x_3^F$. We then add $y_4^T$, representing a true output and $y_{1 \rightarrow 4}^F$, $y_{2 \rightarrow 4}^F$, and $y_{3 \rightarrow 4}^F$, representing false outputs. The rules $x_1^T + y_{1 \rightarrow 4}^F \rightarrow \emptyset$, $x_2^T + y_{2 \rightarrow 4}^F \rightarrow \emptyset$, and $x_3^F + x_1^T \rightarrow \emptyset$ are then applied to the system, removing all reactant species in these rules. The remaining species is $y_{3 \rightarrow 4}^F$, indicating a false output. 

\parae{OR Gate.} To compute an OR gate, a single false output species $y_i^F$ and $|f_i^{in}|$ copies of the true output species $y_{j \rightarrow i}^T$ are added in. In one step, the false input and true output species  delete each other. Additionally, if at least one true input species exists, it deletes the lone false output species along with itself, guaranteeing a true output. The only output species remaining after the step are those whose truth value matches the intended output.



\parae{NOT Gate.} To compute a NOT gate, a single copy of the true and false output species are added in. In a single step, the input and output species that share the same truth value $b$ delete each other, leaving the complement of the input species as the remaining output species.

\begin{table}[t]
\begin{center}
\begin{tabular}{ | c | c c | c | c | c c | c |}\hline

\textbf{Gate Type} & \multicolumn{2}{|c|}{\textbf{Step}} & \textbf{Relevant Rules} & \textbf{Description} \\ \hline




\textbf{AND} & \textit{Add} & \makecell{$y_{i}^T$ \\ $\forall j \in f_i^{in}:$ $y_{j \rightarrow i}^F$} & \makecell{$x_j^T + y_{j \rightarrow i}^F \rightarrow \emptyset$ \\ $x_j^F + y_i^T \rightarrow \emptyset$} & \makecell{An input species with a certain \\ truth value deletes the \\ complement output species.} \\ \hline

 \textbf{OR} & \textit{Add} & \makecell{$y_{i}^F$ \\ $\forall j \in f_i^{in}:$ $y_{j \rightarrow i}^T$} & \makecell{$x_j^T + y_i^F \rightarrow \emptyset$ \\ $x_j^F + y_{j \rightarrow i}^T \rightarrow \emptyset$} & \makecell{An input species with a certain \\ truth value deletes the \\ complement output species.} \\ \hline

 \textbf{NOT} & \textit{Add} & \makecell{$y_{i}^T$ \\ $y_{i}^F$} & \makecell{$x_j^T + y_i^T \rightarrow \emptyset$ \\ $x_j^F + y_i^F \rightarrow \emptyset$} & \makecell{The input and output species that \\ share the same truth value delete \\ each other.} \\ \hline

\end{tabular}
\caption{\textit{(2, 0) rules and steps for computing AND, OR, and NOT gates.}}\label{tab:(2,0)_AND_OR_NOT}
\end{center}
\end{table}


\parae{Majority Gate.} 
To compute a majority gate, all input species are first converted into a new species $a_i^b$ (Step 1). These species retain the same index and truth value of their original inputs. If the number of bits inputted into a majority gate is even, then an extra \emph{false} input species should be added in. The species $b_i^b$ is then added (Step 2). This species performs the majority operation across all $a_i^b$ species. Any species that represents the minority inputs are deleted. The remaining species are then converted into the matching output species (Step 3). 

\begin{table}[t]
    \centering
    \begin{tabular}{| c | c  c | c | c |}\hline
        \multicolumn{3}{|c|}{\textbf{Steps}} & \textbf{Relevant Rules} & \textbf{Description} \\ \hline
        
        \textbf{1} & \textit{Add} & \makecell{$|f_i^{in}| \cdot a_i^T$ \\ $|f_i^{in}| \cdot a_i^F$} 
        & \makecell{\vspace{-10pt} \\ $\forall j \in f_i^{in}:$ \\ $x_j^T + a_i^F \rightarrow \emptyset$ \\ $x_j^F + a_i^T \rightarrow \emptyset$ \\ \vspace{-9pt}}
        & \makecell{$\forall j \in f_i^{in}$, convert $x_{j}^b$ input \\ species into $a_i^b$ species.} \\ \hline
        \textbf{2} & \textit{Add} & \makecell{$\lfloor |f_i^{in}|/2 \rfloor \cdot b_i^T$ \\ $\lfloor |f_i^{in}|/2 \rfloor \cdot b_i^F$} 
        & \makecell{$a_i^T + b_i^F \rightarrow \emptyset$ \\ $a_i^F + b_i^T \rightarrow \emptyset$}
        & \makecell{Adding $\lfloor |f_i^{in}|/2 \rfloor$ amounts of $b_i^T$ and \\ $b_i^F$ species will delete all of the \\ minority species, leaving some amount \\ of the majority species remaining.} \\ \hline
        \textbf{3} & \textit{Add} & \makecell{\vspace{-8pt} \\ $y_i^T$ \\ $y_i^F$ \\ \vspace{-10pt}} 
        & \makecell{$a_i^T + y_i^F \rightarrow \emptyset$ \\ $a_i^F + y_i^T \rightarrow \emptyset$}
        & \makecell{Convert $a_i^b$ into the proper output \\ species ($y_i^b$).} \\ \hline
    \end{tabular}
    \caption{\textit{(2, 0) rules and steps for computing majority gates.}
    }\label{tab:(2,0)_MAJ}
\end{table}



    

\subsection{Formula Computation Example}\label{2_0_example}

We demonstrate a simple example of computing a threshold formula under our constructions. The formula is the same as in Figure \ref{fig:indexingformula}. It has four inputs: $x_1$, $x_2$, $x_3$, and $x_4$. At the first depth level, $x_1$ and $x_2$ fan into an AND gate, as does $x_3$ and $x_4$. Both gate outputs are then fanned into an OR gate, whose output represents the final value of the formula. 

The initial configuration consists of bit species that correlate to the input values for the formula. Step 1 converts the species into input species for the first depth level. Step 2 then performs all gate operations at the first depth level. Step 3 converts the output species of the gates into input species for the next and final depth level. Step 4 computes the only gate at this level. Finally, Step 5 coverts that gate's output into an input species that represents the final output of the formula. A more detailed explanation of computing the formula is in Table \ref{tab:(2,0)_formula_example}.

\begin{table}[t]
    
    \begin{subtable}{.6\linewidth}
        \renewcommand{\arraystretch}{.5}
        \centering
        \begin{tabular}{| c | c  c | c |}\hline
            \multicolumn{4}{|c|}{\textbf{Initial Configuration: $y_1^T$, $y_2^F$, $y_3^F$, $y_4^T$}} \\ \hline
            \multicolumn{3}{|c|}{\textbf{Step}} & \textbf{Relevant Rules} \\ \hline
            \textbf{1} & \textit{Add} & \makecell{$x_1^T$, $x_2^T$, $x_3^T$, $x_4^T$ \\ $x_1^F$, $x_2^F$, $x_3^F$, $x_4^F$} 
            & \makecell{\vspace{-10pt}
            \\ $y_1^T + x_1^F \rightarrow \emptyset$ \\ $y_2^F + x_2^T \rightarrow \emptyset$ \\ $y_3^F + x_3^T \rightarrow \emptyset$ \\ $y_4^T + x_4^F \rightarrow \emptyset$ \\ \vspace{-9pt}
            } \\ \hline
            \textbf{2} & \textit{Add} & \makecell{$y_5^T$, $y_{1 \rightarrow 5}^F$, $y_{2 \rightarrow 5}^F$ \\ $y_6^T$, $y_{3 \rightarrow 6}^F$, $y_{4 \rightarrow 6}^F$} 
            & \makecell{$x_1^T + y_{1 \rightarrow 5}^F \rightarrow \emptyset$ \\ $x_2^F + y_5^T \rightarrow \emptyset$ \\ $x_3^F + y_6^T \rightarrow \emptyset$ \\ $x_4^T + y_{4 \rightarrow 6}^F \rightarrow \emptyset$} \\ \hline
            \textbf{3} & \textit{Add} & \makecell{$x_5^T$, $x_6^T$ \\ $x_5^F$, $x_6^F$} 
            & \makecell{\vspace{-10pt}
            \\ $y_{2 \rightarrow 5}^F + x_5^T \rightarrow \emptyset$ \\ $y_{3 \rightarrow 6}^F + x_6^T \rightarrow \emptyset$ \\ \vspace{-9pt}
            } \\ \hline
            \textbf{4} & \textit{Add} & \makecell{$y_{5 \rightarrow 7}^T$, $y_{6 \rightarrow 7}^T$, $y_7^F$} 
            & \makecell{\vspace{-10pt} 
            \\ $y_{5 \rightarrow 7}^F + x_5^T \rightarrow \emptyset$ \\ $y_{6 \rightarrow 7}^F + x_6^T \rightarrow \emptyset$ \\ \vspace{-9pt}
            } \\ \hline
            \textbf{5} & \textit{Add}
            & \makecell{\vspace{-8pt}
            \\ $x_7^T$, $x_7^F$ \\ \vspace{-10pt}
            }
            & \makecell{$y_7^F + x_7^T \rightarrow \emptyset$} \\ \hline
        \end{tabular}
        
        \caption{
        }\label{tab:(2,0)_formula_example}
    \end{subtable}%
    \begin{subtable}{.4\linewidth}
        \renewcommand{\arraystretch}{1.}
        \centering
        \begin{tabular}{| c  c  | c |}\hline
            
            \multicolumn{2}{|c|}{\textbf{Step}} & \textbf{Relevant Rules}  \\ \hline
            
            
            
            
            
            
            
            \textit{Add} & \makecell{$x_i^T$ \\ $x_i^F$ \\} 
            
            & \makecell{\vspace{-8pt} \\ $y_i^T + x_i^F \rightarrow \emptyset$ \\ $y_i^F + x_i^T \rightarrow \emptyset$ \\ $y_{j \rightarrow i}^T + x_i^F \rightarrow \emptyset$ \\ $y_{j \rightarrow i}^F + x_i^T \rightarrow \emptyset$ \\ \vspace{-8pt}}
            
             \\\hline
            
        \end{tabular}
        \caption{}\label{tab:(2,0)_convert}
    \end{subtable} 
    \caption{(a) (2, 0) rules and steps for computing the formula in Figure \ref{fig:indexingformula}. (b) (2, 0) rules and step for converting outputs to inputs per depth level. Add species for that represent true and false inputs. Delete the species that are the complement of the output. Only the correct input species remains.
    }
\end{table}


\subsection{Threshold Formula Computation}\label{2_0_computation}
We now introduce another step with rules that convert the output species of one depth level into input species for the next level, enabling the complete computation of a TF. We then prove the computational complexity of computing TFs within our system.

\parae{Depth Traversal.}
To enable the traversal of every gates' bits at a specific depth level to the next level, every output species is converted into an input species in one step. The same truth value and index is retained between the output and input species. Table \ref{tab:(2,0)_convert} shows how to compute depth level traversal for an output bit with index $i$.

    
    
    
    
    
    
    
    
    
    
    

\begin{theorem}\label{(2,0)_formula}
Threshold formulas (TF) can be computed with multiple-input relaxation by a step CRN with only $(2,0)$ rules with upper bounds of $O(G)$ species, $O(D)$ steps, and $O(G)$ volume.
\end{theorem}
\begin{proof}
Each gate of a TF is represented by a constant number of species, resulting in $O(G)$ unique species.
All gates at a given depth level are computed simultaneously in constant steps. Computing a TF therefore requires $O(D)$ steps.

It is possible not all species that are no longer needed after computing a specific gate are deleted. For example, computing an AND gate with three false inputs leaves two of those species in the configuration. While this does not cause computation errors, the volume will increase. Therefore, it is possible for only a fraction of the $O(G)$ species added throughout the simulation to be deleted, resulting in $O(G)$ volume.
\end{proof}
\section{Computation of TCs with Exponential Volume}\label{2_0_circuits}

\begin{figure}[t]
    \centering
    \begin{subfigure}[t] {0.4\textwidth}
        \centering
        \includegraphics[width=.9\textwidth]{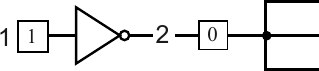}
        \caption{}
        \label{fig:unboundedfanout}
    \end{subfigure}
    \begin{subfigure}[t] {0.5\textwidth}
        \centering
        \includegraphics[width=.9\textwidth]{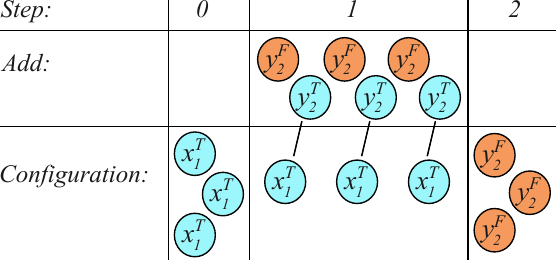}
        \caption{}
        \label{fig:fanoutsteps}
    \end{subfigure}
    \caption{(a) A NOT gate with a fan-out of three. (b) Computing the NOT gate in Figure \ref{fig:unboundedfanout} in (2, 0) rules.
    }\label{fig:sthsnth}
\end{figure}

In this section, we slightly alter the approach presented in Section \ref{2_0_formulas} to enable computation of Threshold Circuits $(TC)$. We show in Section \ref{2_0_exp_vol} how modifying our volume to be exponentially-sized allows the system to account for unbounded fan-out outside of the first depth level, enabling the computation of TCs. Theorem \ref{(2,0)_circuit} shows our system computes TCs with $O(G)$ species, $O(D)$ steps, and $O(G {F_{out}}^D)$ volume.

\subsection{Threshold Circuit Computation}\label{2_0_exp_vol}

Here, we demonstrate how to account for unbounded fan-out when computing TCs, and show the computational complexity of computing TCs with our system.

\parae{Bits and Gates.}
Section \ref{2_0_bits} shows how input bits, output bits, and indexing are represented. Individual gates and depth traversal are computed using the same methods shown in Sections \ref{2_0_gates} and \ref{2_0_computation}, respectively.

\parae{Unbounded Fan-Out.}
To allow the output of a gate at index $i$ to fan-out $k > 1$, such as in the NOT gate shown in Figure \ref{fig:unboundedfanout}, the count of the species added for all the gates whose output eventually fans into gate $i$ should be multiplied by $k$. The gate computes all input species concurrently with each other, and result in the gate's output species being equivalent in quantity to the fan-out. Figure \ref{fig:fanoutsteps} shows an example of this process for a NOT gate with a fan-out of three.

\parae{Unbounded Fan-Out Example.} Consider an AND gate with a fan-in and fan-out of two. Let the two input bits be true and false. To compute the gate with the correct amount of fan-out, our system's initial configuration consists of two copies of each input species ($x_1^T$, $x_1^T$, $x_2^F$, and $x_2^F$). We then apply the relevant rules to the configuration. Afterwards, we are left with two copies of $y_{2 \rightarrow 1}^F$.

 \begin{theorem}\label{(2,0)_circuit}
 Threshold circuits (TC) can be computed with multiple-input relaxation by a step CRN with only (2, 0) rules with upper bounds of $O(G)$ species, $O(D)$ steps, and $O(G {F_{out}}^D)$ volume.
 \end{theorem}

\section{Exponential Volume Lower Bound for Gate-Wise Simulation}\label{2_0_lower_bound}

In this section, we derive an exponential lower bound for the volume of a step CRN with $(2,0)$ rules that simulates boolean circuits of depth $D$. Our lower bound almost matches the upper bound. 

To prove the lower bound, we design a circuit that is able to be simulated by any step CRN using only $(2,0)$ rules as follows. The circuit has $D$ stages such that each stage of the circuit has $O(1)$ layers. 
We establish a recursive inequality for the CRN volume over three consecutive stages which implies an exponential lower bound for species in the input stage. We show a proof that the lower bound of volume in a step CRN that 
uses gate-wise simulation to simulate a boolean circuit with only $(2,0)$ rules is $2^{\Omega(D)}$. 




\begin{definition}\label{natural-simulation-def}
A step CRN uses \emph{gate-wise simulation} to simulate a circuit $V(\cdot)$
if each gate $g$ is assigned $c_{1,g}$ copies of species $1_g$, which represents output $1$, and $c_{0, g}$ copies of $0_g$, which represents output $0$. When gate $g$ computes an output, it will be either $c_{1,g}$ copies of species $1_g$ for the case $1$, or $c_{0,g}$ copies of species $0_g$ for the case $0$. We define $C(g)=c_{1,g}+c_{0,g}$ to be the number of species used for gate $g$.
There is a special case that $g$ is one of the input bits (source gates with fan-in 0) that satisfies $c_{1,g}=0$ or $c_{0,g}=0$, as each input bit is either $0$ or $1$. Let a step CRN have $k$ steps $0,1,\cdots, k-1$ as defined in Section~\ref{def-step-CRN-section}. It also satisfies the conditions:
\begin{itemize} 
    \item Every gate $g$ enters its complete state at exactly one step $i$, which is denoted by $complete(g)=i$. After step $i$, the system releases $c_{a,g}$ copies of species $a_g$ and removes all copies of species $(1-a)_g$ to represent gate $g$ having the output $a$. After step $complete(g)$, the system does not generate any additional copy of $1_g$ or $0_g$ (it may keep some existing copies of $a_g$). The output of gate $g$ determines the simulation according to the logic of circuit.
    \item For two different gates $g_1$ and $g_2$, if there is a directed path from $g_1$ to $g_2$ ($g_1$'s output may affect $g_2$'s output)  in the circuit $V(\cdot)$, then $complete(g_1)<complete(g_2)$.
\end{itemize}

If a circuit $V(\cdot)$ computes a function $f(x_1,\cdots, x_n)=y_1,\cdots, y_m$ ($\{0,1\}^n\rightarrow \{0,1\}^m$)  and $V(\cdot)$ is simulated using gate-wise simulation in a step CRN, 
define $1_{f(\cdot),i}$ to represent the case $y_i$ to be $1$ and $0_{f(\cdot),i}$ to represent the case $y_i$ to be $0$.
  \end{definition}

By Definition~\ref{natural-simulation-def}, when gate $g$ outputs $1$, all the $c_{0,g}$ copies of $0_g$ are removed and it has $c_{1,g}$ copies of $1_{g}$ to enter the next layer of a circuit. 
The step CRN given 
in Section~\ref{2_0_circuits} uses gate-wise simulation to simulate threshold circuits. Our lower bound result in this section shows that the exponential volume is required.


\begin{lemma}\label{basic-lemma}
Assume that  a step CRN with $(2,0)$ rules simulates a circuit $V(\cdot)$. Let $b\in \{1_g, 0_g\}$ be the output species. If one copy of a species is removed or added, it results in at most one difference in the number of copies of species $b$. 
\end{lemma}
\begin{proof}
   We prove via induction over the total start volume. If the start volume is zero, it is trivial. Adding one copy of species $u$ results in no additional changes since there is only one copy of a species and every rule requires two copies. This is why it affects at most one copy of a species in the output side.

   Consider the case that the start volume is one. We only have one copy of a species $v$.
   If one copy of $u$ is added, there are two possible cases: 1) $u$ and $v$ do not react (no rule $u+v\rightarrow\emptyset$ exists), and 2) $u$ and $v$ react ($u+v\rightarrow\emptyset$). In case 1), the output of the CRN increases by at most one copy. In case 2), the output of the CRN removes at most one copy. 
  
  Assume it is true when the start volume is at most $n$. Consider the start volume to be $n+1$. Let species $u$ lose one copy. Let $u+v \rightarrow \emptyset$ be applied (otherwise, it is trivial). We transform the problem into adding one copy of $v$ into a step CRN with volume $n-1$. This case is proved by inductive hypothesis. The case of adding one copy of $u$ is similar.  
\end{proof}

\begin{lemma}\label{fundamental-lemma}
Let $f(x_1,\cdots, x_n)=y_1, \cdots, y_m$ be a function $\{0,1\}^n\rightarrow \{0,1\}^m$ such that each variable $x_i$ and $y_j$ has a value in $\{0,1\}$. 
If a step CRN with $(2,0)$ rules computes $f(\cdot)$, and changing variable $x_j$ to $1-x_j$ will change $y_{i_1},\cdots, y_{i_t}$ to $1-y_{i_1},\cdots, 1-y_{i_t}$, respectively, then $C(x_j)\ge \sum_{k=1}^t C(y_{i_k})$.
\end{lemma}

\begin{proof}
Assume that there is a step CRN to simulate $f(\cdot)$.
Assume $C(x_j)< \sum_{k=1}^t C(y_{i_k})$.
For each Boolean variable $x$, we also use $x$ to represent the gate that outputs the value of $x$.

When one copy is removed from (or added to) the input side, it may add or decrease at most one copy on the output side by Lemma~\ref{basic-lemma}.
If the output side needs to change its value from $a\in \{0,1\}$ to $1-a$, then it must remove all the copies that represent the current $a$ and add the new copies that represent the value $1-a$.
Assume that the current $x_j$ has value $a\in\{0,1\}$. Consider the transitions of two phases:

\begin{itemize}
    \item Phase 1: Remove $c_{a, x_j}$ copies of $x_j$.
    After $x_j$ is removed, it affects at most $c_{a, x_j}$ copies of species $y_{i_1},\cdots, y_{i_t}$ by Lemma~\ref{basic-lemma}.
    \item Phase 2: Add $c_{1-a, x_j}$ copies of $x_j$ with value $1-a$ over the input side. 
    After adding $c_{1-a, x_j}$ copies of $x_j$ (with flipped value), it affects at most $c_{1-a, x_j}$ copies of species $y_{i_1},\cdots, y_{y_t}$ by Lemma~\ref{basic-lemma}.
\end{itemize}

The output side destroys the old copies of the species for $y_{i_1},\cdots, y_{i_t}$, and adds new copies for $y_{i_1},\cdots, y_{i_t}$. However, the 2 phases affect at most $C(x_j)=c_{a, x_j}+c_{1-a, x_j}< \sum_{k=1}^t C(y_{i_k})$ copies on the output side. 
Thus, a contradiction.
\end{proof}

\begin{definition}
A list of Boolean circuits $\{H_n(\cdot)\}$ is uniform if there is a Turing machine $M(\cdot)$ such that each $H_n(\cdot)$ can be generated by $M(1^n)$ in a polynomial $p(n)$ steps.    
\end{definition}

\begin{theorem}\label{(2,0)-lower-thm} There exist uniform Boolean circuits $\{V_D(x_1, x_2, x_3)\}_{D=1}^{+\infty}$ with each $V_D(x_1, x_2, x_3):\{0,1\}^3\rightarrow \{0,1\}^3$ s.t. each $V_D(x_1, x_2, x_3)$ has depth $O(D)$, size $O(D)$, 3 output bits, and requires $C(g)=2^{\Omega(D)}$ for at least one input gate $g$ 
in a step CRN using gate-wise simulation to simulate $V_D(\cdot)$ with $(2,0)$ rules.
\end{theorem}

\begin{proof}
We construct a circuit that has $O(D)$ layers. It is built in $D$ stages. Each stage has a circuit of depth $O(1)$ to compute function $s(x_1, x_2, x_3)=y_1y_2y_3$.
The function $s(\cdot)$ has the properties: 
\vspace{-.2cm}


\begin{eqnarray}
    s(1,1,1)&=&111, \label{x1x2x3-ineqn}\\
    s(0,1,1)&=&000, \label{x1-y4-ineqn} \\
    s(1,0,1)&=&011, \label{x2-y1-ineqn}\\
    s(1,1,0)&=&101, \label{x3-y2-ineqn}\\
    s(0,0,0)&=&110.
\end{eqnarray}


Define function $s^{(1)}(x_1, x_2, x_3)=s(x_1, x_2, x_3)$ and $s^{(k+1)}(x_1, x_2, x_3)=s(s^{(k)}(x_1, x_2, x_3))$ for all integers $k>1$. The circuit $V(x_1, x_2, x_3)=s^{(D)}(x_1, x_2, x_3)$. We can also represent the circuit $V(x_1, x_2, x_3)=s^{(D)}(x_1, x_2, x_3)=s_{D-1}\circ s_{D-2}\circ \cdots \circ s_0(x_1, x_2, x_3)$, where $s_i(\cdot)$ represent the function $s(\cdot)$ at stage $i$. The circuit $V(x_1, x_2, x_3)$, which computes $s^{(D)}(x_1,x_2, x_3)$ links the $D$ circuits $V_s(x_1, x_2, x_3)$ that compute the function $s(x_1, x_2, x_3)$.
The three output bits for $s_k(\cdot)$ at stage $k$ become three input bits of $s_{k-1}(\cdot)$ at stage $k-1$.

Let the output of the circuit be stage $0$. The input stage has the largest stage index.
Consider the general case. Let $C_k(u)$ be the number of copies of species $u$ in stage $k$. If $u$ is computed by a gate $g$, we let $C_k(u)=C(g)$. Let $v_{i, k}$ be the variable $v_i$ at stage $k$. As we have three output bits $y_{1,0}, y_{2,0}, y_{3,0}$ in the last layer (layer $D$), each bit $y_{i,0}$ must have a copy of species to represent its $0,1$-value (see Definition~\ref{natural-simulation-def}).  Thus,
 
 
\begin{eqnarray}
C_D(y_{1,0})\ge 1, \ C_D(y_{2,0})\ge 1, \ C_D(y_{3,0})\ge 1.\label{initial-yyy-ineqn}   
\end{eqnarray}
 
When $x_1x_2x_3$ is changed from $111$ to $011$ (by flipping $x_1$), the output $y_1y_2y_3$ is changed from $111$ to $000$.
By Equations (\ref{x1x2x3-ineqn}) and (\ref{x1-y4-ineqn}) and Lemma~\ref{fundamental-lemma},
 we have 
 \begin{eqnarray}
 C_k(x_{1,k})\ge  C_k(y_{1,k})+C_k(y_{2,k})+C_k(y_{3,k}).  \label{x1-y1-y2-y3-ineqn}   
 \end{eqnarray}


When $x_1x_2x_3$ is changed from $111$ to $101$ (by flipping $x_2$), the output $y_1y_2y_3$ is changed from $111$ to $011$.
 By Equations (\ref{x1x2x3-ineqn}) and (\ref{x2-y1-ineqn})  and Lemma~\ref{fundamental-lemma}, we have Equation \ref{x-to-y-ineqn} below. When $x_1x_2x_3$ is changed from $111$ to $110$ (by flipping $x_3$), the output $y_1y_2y_3$ is changed from $111$ to $101$.
  By Equations (\ref{x1x2x3-ineqn}) and (\ref{x3-y2-ineqn})  and Lemma~\ref{fundamental-lemma}, we have Equation \ref{unlabelledeq} below.


\begin{eqnarray}
 C_k(x_{2,k})\ge C_k(y_{1,k}).  \label{x-to-y-ineqn} \\
 C_k(x_{3,k})\ge C_k(y_{2,k}),  \label{unlabelledeq}
\end{eqnarray}

 

When $y_{i, k}$ is equal to $x_{i, k-1}$ as the output of $s_{k}(\cdot)$ becomes the input of $s_{k-1}(\cdot)$.
We have 
\begin{eqnarray}
C_k(y_{i, k})\ge C_{k-1}(x_{i, k-1})\ \  for\ i=1,2,3. \label{yk-xk-1-ineqn}
\end{eqnarray}

In each stage, the input to the function $s(\cdot)$ can reach all cases $000,011,101,$ $110,$$111$
by adjusting the 3 input bits of the circuit. When the input is $111$, the function $s(\cdot)$
gives the same output $111$
at all phases.
Through a simple repetition of the above inequalities, we derive a $2^{\Omega(D)}$ volume lower bound.

\begin{eqnarray}
 C_k(x_{1, k})  &\ge& C_{k-1}(x_{1, k-1})+C_{k-1}(x_{2, k-1})+C_{k-1}(x_{3, k-1}) (by\ (\ref{x1-y1-y2-y3-ineqn}), (\ref{yk-xk-1-ineqn}))\\
 &\ge& C_{k-1}(x_{1, k-1})+C_{k-1}(y_{1, k-1})\ \ (by\ inequality\ (\ref{x-to-y-ineqn}))\\
  &\ge& C_{k-1}(x_{1, k-1})+C_{k-2}(x_{1, k-2}). \ \ (by\ inequality\ (\ref{yk-xk-1-ineqn})) \label{last-ineqn}
\end{eqnarray}

Let $a_0, a_1,\cdots$ be the Fibonacci series with $a_0=a_1=1$ and recursion $a_{k}=a_{k-1}+a_{k-2}$ for all $k>1$. By inequalities (\ref{initial-yyy-ineqn}), (\ref{x1-y1-y2-y3-ineqn}), and the fact that every input bit affects the output bit in $s(\cdot)$, we have $C_0(x_{1, 0})\ge 1$ and $C_1(x_{1,1})\ge 1$. This is because when the three input bits are $111$, we need bit $x_1$ to make the output bits $111$. 
By inequality (\ref{last-ineqn}), we have $C_{k}(x_{1,k})\ge a_k$ for all $k\ge 0$.  
\end{proof}

\section{Conclusions and Open Problems}


In this paper we show how bimolecular void rules, a subset of reaction rules with low power compared to traditional CRNs, become capable of computing threshold formulas and circuits in the step CRN model under gate-wise simulation. We also prove that simulating circuits under this technique requires an exponential lower bound volume that matches the upper bound of our construction methods.

These results naturally lead to some promising future research directions. One approach is constructing another method for simulating threshold circuits under only (2,0) rules. A more general simulation technique could have the benefit of computing circuits without the exponential-sized volume gate-wise simulation requires.
Furthermore, our step CRN definition requires the system to reach a terminal configuration before moving to the next step. Relaxing this definition so that a system may reach a step without entering a terminal configuration can make the model more valuable to general CRNs, where reachability to a terminal configuration is not guaranteed.

\bibliographystyle{plainurl}
\bibliography{20crn}

\begin{thebibliography}{10}

\bibitem{Alaniz:2022:ARXIV}
Robert~M. Alaniz, Bin Fu, Timothy Gomez, Elise Grizzell, Andrew Rodriguez,
  Robert Schweller, and Tim Wylie.
\newblock Reachability in restricted chemical reaction networks, 2022.
\newblock arXiv:2211.12603.
\newblock \href {http://arxiv.org/abs/2211.12603} {\path{arXiv:2211.12603}}.

\bibitem{anderson2024computing}
Rachel Anderson, Alberto Avila, Bin Fu, Timothy Gomez, Elise Grizzell, Aiden
  Massie, Gourab Mukhopadhyay, Adrian Salinas, Robert Schweller, Evan Tomai,
  and Tim Wylie.
\newblock Computing threshold circuits with void reactions in step chemical
  reaction networks, 2024.
\newblock arXiv:2402.08220.
\newblock \href {http://arxiv.org/abs/2402.08220} {\path{arXiv:2402.08220}}.

\bibitem{Angluin:2006:DC}
Dana Angluin, James Aspnes, Zo\"{e} Diamadi, Michael~J. Fischer, and Ren\'{e}
  Peralta.
\newblock Computation in networks of passively mobile finite-state sensors.
\newblock {\em Distribed Computing}, 18(4):235–253, mar 2006.

\bibitem{angluin2008simple}
Dana Angluin, James Aspnes, and David Eisenstat.
\newblock A simple population protocol for fast robust approximate majority.
\newblock {\em Distributed Computing}, 21:87--102, 2008.

\bibitem{angluin2007computational}
Dana Angluin, James Aspnes, David Eisenstat, and Eric Ruppert.
\newblock The computational power of population protocols.
\newblock {\em Distributed Computing}, 2007.

\bibitem{Aris:1965:ARMA}
Rutherford {Aris}.
\newblock Prolegomena to the rational analysis of systems of chemical
  reactions.
\newblock {\em Rational Mechanics and Analysis}, 19(2):81--99, jan 1965.

\bibitem{Aris:1968:ARMA}
Rutherford {Aris}.
\newblock Prolegomena to the rational analysis of systems of chemical reactions
  ii. some addenda.
\newblock {\em Rational Mechanics and Analysis}, 27(5):356--364, jan 1968.

\bibitem{arkin1994computational}
Adam Arkin and John Ross.
\newblock Computational functions in biochemical reaction networks.
\newblock {\em Biophysical journal}, 67(2):560--578, 1994.

\bibitem{beiki2018real}
Z~Beiki, Z~Zare Dorabi, and Ali Jahanian.
\newblock Real parallel and constant delay logic circuit design methodology
  based on the dna model-of-computation.
\newblock {\em Microprocessors and Microsystems}, 61:217--226, 2018.

\bibitem{cardelli2012cell}
Luca Cardelli and Attila Csik{\'a}sz-Nagy.
\newblock The cell cycle switch computes approximate majority.
\newblock {\em Scientific reports}, 2(1):656, 2012.

\bibitem{cardelli2018chemical}
Luca Cardelli, Marta Kwiatkowska, and Max Whitby.
\newblock Chemical reaction network designs for asynchronous logic circuits.
\newblock {\em Natural computing}, 17:109--130, 2018.

\bibitem{chen2014deterministic}
Ho-Lin Chen, David Doty, and David Soloveichik.
\newblock Deterministic function computation with chemical reaction networks.
\newblock {\em Natural computing}, 13(4):517--534, 2014.

\bibitem{Cook:2009:AB}
Matthew Cook, David Soloveichik, Erik Winfree, and Jehoshua Bruck.
\newblock {\em Algorithmic Bioprocesses}, chapter Programmability of Chemical
  Reaction Networks, pages 543--584.
\newblock Springer, 2009.

\bibitem{dalchau2015probabilistic}
Neil Dalchau, Harish Chandran, Nikhil Gopalkrishnan, Andrew Phillips, and John
  Reif.
\newblock Probabilistic analysis of localized dna hybridization circuits.
\newblock {\em ACS synthetic biology}, 4(8):898--913, 2015.

\bibitem{ellis2019robust}
Samuel~J Ellis, Titus~H Klinge, and James~I Lathrop.
\newblock Robust chemical circuits.
\newblock {\em Biosystems}, 186:103983, 2019.

\bibitem{eshra2013odd}
Abeer Eshra and Ayman El-Sayed.
\newblock An odd parity checker prototype using dnazyme finite state machine.
\newblock {\em IEEE/ACM Transactions on Computational Biology and
  Bioinformatics}, 11(2):316--324, 2013.

\bibitem{Fan_Fan_Wang_Dong_2018}
Daoqing Fan, Yongchao Fan, Erkang Wang, and Shaojun Dong.
\newblock A simple, label-free, electrochemical dna parity generator/checker
  for error detection during data transmission based on
  “aptamer-nanoclaw”-modulated protein steric hindrance.
\newblock {\em Chemical Science}, 9(34):6981–6987, 2018.
\newblock \href {https://doi.org/10.1039/C8SC02482K}
  {\path{doi:10.1039/C8SC02482K}}.

\bibitem{fan2022engineering}
Daoqing Fan, Jun Wang, Jiawen Han, Erkang Wang, and Shaojun Dong.
\newblock Engineering dna logic systems with non-canonical dna-nanostructures:
  Basic principles, recent developments and bio-applications.
\newblock {\em Science China Chemistry}, 65(2):284--297, 2022.

\bibitem{jiang2013digital}
Hua Jiang, Marc~D Riedel, and Keshab~K Parhi.
\newblock Digital logic with molecular reactions.
\newblock In {\em Intl. Conf. on Computer-Aided Design (ICCAD)}, ICCAD'13,
  pages 721--727, 2013.

\bibitem{Karp:1969:JCSS}
Richard~M. Karp and Raymond~E. Miller.
\newblock Parallel program schemata.
\newblock {\em Journal of Computer and System Sciences}, 3(2):147--195, 1969.

\bibitem{lin2020mining}
Yu-Chou Lin and Jie-Hong~R Jiang.
\newblock Mining biochemical circuits from enzyme databases via boolean
  reasoning.
\newblock In {\em 39th Intl. Conf. on Computer-Aided Design}, pages 1--9, 2020.

\bibitem{magri2009fluorescent}
David~C Magri.
\newblock A fluorescent and logic gate driven by electrons and protons.
\newblock {\em New Journal of Chemistry}, 33(3):457--461, 2009.

\bibitem{Mailloux_Guz_Zakharchenko_Minko_Katz_2014}
Shay Mailloux, Nataliia Guz, Andrey Zakharchenko, Sergiy Minko, and Evgeny
  Katz.
\newblock Majority and minority gates realized in enzyme-biocatalyzed systems
  integrated with logic networks and interfaced with bioelectronic systems.
\newblock {\em The Journal of Physical Chemistry B}, 118(24):6775–6784, Jun
  2014.
\newblock \href {https://doi.org/10.1021/jp504057u}
  {\path{doi:10.1021/jp504057u}}.

\bibitem{Petri:1962:PHD}
Carl~Adam Petri.
\newblock {\em Kommunikation mit Automaten}.
\newblock PhD thesis, Rheinisch-Westf{\"a}lischen Institutes f{\"u}r
  Instrumentelle Mathematik an der Universit{\"a}t Bonn, 1962.

\bibitem{qian2011scaling}
Lulu Qian and Erik Winfree.
\newblock Scaling up digital circuit computation with dna strand displacement
  cascades.
\newblock {\em science}, 332(6034):1196--1201, 2011.

\bibitem{Qian_Winfree_2011}
Lulu Qian and Erik Winfree.
\newblock A simple dna gate motif for synthesizing large-scale circuits.
\newblock {\em Journal of The Royal Society Interface}, 8(62):1281–1297, Feb
  2011.

\bibitem{soloveichik2008computation}
David Soloveichik, Matthew Cook, Erik Winfree, and Jehoshua Bruck.
\newblock Computation with finite stochastic chemical reaction networks.
\newblock {\em natural computing}, 7(4):615--633, 2008.

\bibitem{thachuk2015leakless}
Chris Thachuk, Erik Winfree, and David Soloveichik.
\newblock Leakless dna strand displacement systems.
\newblock In {\em 21st International Conference on DNA Computing and Molecular
  Programming}, DNA'15, pages 133--153. Springer, 2015.

\bibitem{winfree2019chemical}
Erik Winfree.
\newblock Chemical reaction networks and stochastic local search.
\newblock In {\em 25th Intl. Conf. on DNA Computing and Molecular Programming},
  DNA'19, pages 1--20, 2019.

\bibitem{Xiao_Zhang_Zhang_Chen_Shi_2020}
Wei Xiao, Xinjian Zhang, Zheng Zhang, Congzhou Chen, and Xiaolong Shi.
\newblock Molecular full adder based on dna strand displacement.
\newblock {\em IEEE Access}, 8:189796–189801, 2020.

\end{thebibliography}

\end{document}